\documentstyle[aps,epsf]{revtex}
\input amssym.def

\begin{document}
\date{\today}
\title{Evolutionary Dynamics While Trapped in Resonance: 
A Keplerian Binary System Perturbed by
Gravitational Radiation}
\author{
C. Chicone\thanks{ Department of  Mathematics,
University of Missouri, Columbia, MO 65211.
Supported by
 the NSF grant  DMS-9531811.} ,\quad
B. Mashhoon\thanks{ Department of  Physics and Astronomy,
University of Missouri, Columbia, MO 65211.
}  \quad
and D. G. Retzloff\thanks{ Department of Chemical Engineering,
University of Missouri, Columbia, MO 65211.}
}

\maketitle

\begin{abstract}
The method of averaging is used to investigate the
phenomenon of capture into resonance for a model that describes
a Keplerian
binary system influenced by radiation damping and external normally
incident periodic gravitational radiation.
The dynamical evolution  of the binary orbit while
trapped in resonance is elucidated using the second order partially averaged
system.
This method provides a theoretical framework that can be used to explain the
main evolutionary dynamics of a physical system that has
been trapped in resonance.
\end{abstract}

\section{Introduction}
In two previous papers~\cite{cmr,cmr1}, we considered the long term nonlinear
evolution of a Keplerian binary system that is perturbed by a normally incident
periodic gravitational wave,  and in a recent work~\cite{cmr2} we considered the
additional effect of radiation damping, 
which is of interest in connection with
the observed behavior of the Hulse-Taylor 
binary pulsar PSR~1913+16~\cite{hulse,taylor}.  
These studies have been concerned with the issue
of {\em gravitational ionization}, i.e. the possibility that an external
periodic gravitational wave could ionize a Keplerian binary system over a long
period of time.  The impetus for this subject has come from the close conceptual
analogy between gravitational ionization and a fundamental physical process,
namely, the electromagnetic ionization of a hydrogen atom.  That is, in these
studies one hopes to learn about the disposition of gravitational radiation to
transfer energy and angular momentum in its interaction with matter.  In our
recent investigation~\cite{cmr2}, we encountered an interesting dynamical
phenomenon connected with the passage of the binary orbit through resonance.  As
the binary system loses energy and angular momentum by emitting gravitational
waves, its semimajor axis and eccentricity decrease monotonically on the
average; however, this process of inward spiraling could stop if the system is
captured into a resonance.  The resonance condition fixes the semimajor axis;
therefore, if the semimajor axis decreases to the resonant value and the orbit
is trapped, it then maintains this value {\it on the average} while the external
perturbation deposits energy into the system to compensate for the radiation
damping {\it on the average}.  
It turns out that along with this energy deposition, the external
tidal perturbation can also deposit angular momentum into the binary orbit so that
its  eccentricity decreases considerably during the passage through resonance.
This was the situation for the particular instance of resonance trapping 
reported in~\cite{cmr2}. 
In general, the orbital angular momentum can increase or decrease while the orbit
is trapped in resonance.
Figure~\ref{fig1} depicts passage through a
$(1:1)$ resonance in which the eccentricity decreases.
A similar phenomenon but with an increase in eccentricity
has been reported in the recent numerical study
of the three-body problem by Melita and Woolfson~\cite{melita}. 
The same situation can occur over a long time-scale in a $(4:1)$ resonance in our
model as depicted in Figure~\ref{fig3}. 
The dynamical phenomena associated with an orbit trapped in a resonance
occur over many Keplerian periods of the osculating ellipse; 
therefore, it is natural to  average the dynamical equations over the 
orbital period at resonance. This partial averaging removes the ``fast'' motion
and allows us to see more clearly the ``slow'' motion during trapping.
It is possible
to provide a description of the slow motion as well as 
a theoretical justification for the transfer of angular momentum by
means of the method of second order partial averaging near a resonance.  The
purpose of the present paper is to study this phenomenon theoretically using the
method of averaging for the resonances that occur in a Keplerian binary that is
perturbed by the emission and absorption of gravitational radiation.

Let us consider the simplest model involving a perturbed Keplerian (i.e.
nonrelativistic) binary that contains the effects of radiation reaction damping
and external tidal perturbation in the lowest (i.e. quadrupole) order.  The
tidal perturbation could be due to external masses and gravitational radiation;
in fact, we choose in what follows a normally incident periodic gravitational
wave for the sake of simplicity \cite{cmr2}.  The Keplerian binary orbit under
the combined perturbations due to the emission and absorption of gravitational
radiation in the quadrupole approximation is given in our model
by the equation of relative
motion
\begin{equation}\label{RelativeMotion}
\frac{d^2x^i}{dt^2} + \frac{kx^i}{r^3} + {\cal R}^i = 
  - \epsilon {\cal K}_{ij}\; x^j,
\end{equation}
where ${\bf x}(t)$ is the relative two-body orbit, $r = |{\bf x}|$, and $k =
G_0(m_1+m_2)$.  The binary consists of two point masses $m_1$ and $m_2$,
$m_1+m_2 = M$, that move according to ${\bf x}_1(t) = (m_2/M)\; {\bf x}(t)$ and
${\bf x}_2(t) = -(m_1/M)\; {\bf x}(t)$
, so that the center of mass of the system is at the origin of the
coordinates.  In fact, the center of mass of the binary can be taken to be at
rest in the approximation under consideration here.  The explicit expressions
for ${\cal R}$ and ${\cal K}$ are
\begin{equation}\label{RadiationReaction}
{\cal R} = \frac{4G_0^2m_1m_2}{5c^5r^3}
 \left[\left(12 v^2-30\dot{r}^2-\frac{4k}{r}\right){\bf v} 
 -\frac{\dot{r}}{r}\left(36v^2-50\dot{r}^2+\frac{4k}{3r}\right){\bf x}\right],
\end{equation}
and
\begin{equation}\label{TidalForce}
{\cal K} = \Omega^2 \left[ \begin{array}{ccc}
\alpha \cos{\Omega t} & \beta \cos{(\Omega t + \rho)} & 0 \\
\beta \cos{(\Omega t + \rho)} & -\alpha \cos{\Omega t} & 0 \\
0 & 0 & 0
\end{array} \right],
\end{equation}
where ${\bf v}=\dot{\bf x}(t)$, an overdot denotes differentiation with respect
to time, $\alpha$ and $\beta$ are of the order of unity and are the amplitudes
of the two independent states of linear polarization of the normally incident wave, $\rho$
is a constant phase, and $\Omega$ is the frequency of the external wave.

It is interesting to transform the dynamical 
system~(\ref{RelativeMotion})--(\ref{TidalForce}) to dimensionless form.  
To accomplish this, let ${\bf x}\rightarrow R_0 {\bf x}$ and $t \rightarrow T_0 t$ 
where $R_0$ and $T_0$ are
scale parameters.  Under this transformation, $k \rightarrow k T_0^2/R_0^3$ and
${\cal K}$ remains unchanged if we let $\Omega \rightarrow \Omega/T_0$.  Let us
further restrict $R_0$ and $T_0$ by the relation $k T_0^2 = R_0^3$, so that we
can set $k=1$ in the dynamical equations; for instance, this condition is
satisfied if $R_0$ is the initial semimajor axis of the unperturbed Keplerian
orbit and $2\pi T_0$ is its period.  Furthermore, 
the dynamical system~(\ref{RelativeMotion})--(\ref{TidalForce}) is planar; 
therefore, it is
convenient to express these dimensionless equations in polar coordinates 
$(r, \theta)$ in the orbital plane. The result is
\begin{eqnarray}\label{MathEQM}
\frac{dr}{dt} & = & p_r, \nonumber \\
\frac{d\theta}{dt} & = & \frac{p_{\theta}}{r^2}, \nonumber \\
\frac{dp_r}{dt} & = & -\frac{1}{r^2} + \frac{p_{\theta}^2}{r^3} +
                 \frac{4\delta p_r}{r^3}\left(p_r^2+6\frac{p_{\theta}^2}{r^2}+
                 \frac{4}{3r}\right) \nonumber \\ 
       & & \hspace*{.25in} -\epsilon r
           \Omega^2 [\alpha\cos{2\theta}\cos{\Omega t}
               +\beta\sin{2\theta}\cos{(\Omega t+\rho)}], \nonumber \\
\frac{dp_{\theta}}{dt} & = & \frac{2\delta p_{\theta}}{r^3}
\left( 9p_r^2 - 6 \frac{p_{\theta}^2}{r^2} + \frac{2}{r}\right) \nonumber \\ 
&& \hspace*{.25in} +\epsilon r^2 \Omega^2 [\alpha\sin{2\theta}\cos{\Omega t}
                -\beta\cos{2\theta}\cos{(\Omega t+\rho)}],
\end{eqnarray}
where $\delta$, $0<\delta<<1$,
is the dimensionless strength of radiation reaction and is given by
\begin{equation}\label{DeltaDefn}
\delta = \frac{4G_0^2m_1m_2}{5c^5T_0R_0},
\end{equation}
while $\epsilon$, $0<\epsilon<<1$, is the dimensionless strength of the
external periodic perturbation.
In this paper, we let $\delta = \epsilon \Delta$, where $\Delta$, 
$0 < \Delta < \infty$, is a parameter that is fixed in the system;  
in this way, we avoid dealing with a two parameter $(\epsilon,\delta)$ 
perturbation problem. In particular, we consider only perturbations that
correspond to fixed directions from the origin of this parameter space.
The full two parameter problem would require the consideration of perturbations
corresponding to all curves in the parameter space. 

In the absence of radiative perturbations $(\epsilon =0$ and $\delta=0)$, the
dynamical system (\ref{MathEQM}) describes a Keplerian ellipse.  It is therefore
useful to
express the dynamical equations (\ref{MathEQM}) in terms of Delaunay variables
that are closely related to the elements of the osculating ellipse.  This is the
ellipse that the relative orbit would describe at time $t$, if the perturbations
were turned off at $t$. 
The osculating ellipse always has the same focus,
which is taken to be the
origin of the (polar) coordinates in the space of relative coordinates.
Let the state of relative motion be described by $({\bf
x}, {\bf v})$, or equivalently $(r, \theta, p_r, p_{\theta})$, at time $t$; then,
the energy of the motion fixes the semimajor axis $a$ of the osculating ellipse
and its eccentricity is subsequently fixed by the orbital angular momentum
$p_{\theta}$.  Only two angles are left to determine the osculating ellipse
completely: the orientation of the ellipse in the orbital plane given by $g$ and
the position on the ellipse measured from the periastron given by the true
anomaly $\hat{v}$.  The latter is obtained from
$p_rp_\theta=e\sin\hat{v}$ and the equation of the
ellipse 
\[ \frac{p_{\theta}^2}{r} = 1+e \cos{\hat{v}}, \] 
and the former from
$\theta - \hat{v}$.  The relevant Delaunay ``action-angle" variables $(L, G,
\ell, g)$ are thus defined by \cite{Kov,Stern}
\begin{eqnarray}
L := a^{1/2}, & \hspace*{.25in} & G := p_{\theta} = L
(1-e^2)^{1/2}, \nonumber \\
\ell := \hat{u} - e \sin{\hat{u}}, & & g := \theta - \hat{v},
\end{eqnarray}
where $\hat{u}$ is the eccentric anomaly of the osculating ellipse, $r =
a(1-e\cos{\hat{u}})$, and $\ell$ is the mean anomaly.  
The dynamical system~(\ref{MathEQM}) in terms of Delaunay variables is given 
briefly in Appendix~\ref{appendixa} and used in the following section.

The Delaunay equations of motion are useful for the investigation of periodic
orbits using the Poincar\'{e} surface of section technique~\cite{poincare}.  It
has been shown in \cite{cmr2} that nearly resonant periodic orbits exist in system
(\ref{MathEQM}) for sufficiently small $\epsilon$ and $\Delta$.  These 
correspond to 
$(m:1)$ resonances, where $(m:n)$ refers to the resonance condition $m\omega =
n\Omega$.  Here $m$ and $n$ are two relatively prime integers and $\omega=1/L^3$ 
is the Keplerian frequency of the orbit.  A linear perturbation treatment
\cite{mashoon1} first revealed resonant absorption at $(m:1)$ resonances.  There
could, in principle, be other periodic orbits whose existence is not revealed by
our method \cite{ccc,cmr}.

In our numerical investigation of the simple nonlinear model described above,
we found~\cite{cmr2} instances of resonance trapping during which the
behavior of the osculating orbit could not be inferred in a simple manner
on the basis of equation~(\ref{MathEQM}). 
However, the dynamics of the {\em averaged} equations
in resonance is simpler to analyze and it turns out that our numerical 
results~\cite{cmr2} can be adequately explained using the second order partially
averaged dynamics. The phenomenon of resonance trapping appears to be
of basic significance for the origin of the solar system; therefore,
it is worthwhile to develop a general theoretical framework for the
study of the evolutionary dynamics while trapped in resonance.

The {\em dynamics} of a system when it is locked in
resonance is interesting in any circumstance involving
more than one degree of freedom; for instance, let us suppose that
the resonance condition
fixes an action variable---say, energy---and for a one dimensional
motion this would then imply that the state of the system at resonance is definite.
However, if other action variables are present, they will not necessarily remain
fixed while the system is trapped in resonance.  
Instead, the state of the
system will in general vary and its dynamics at resonance is best
investigated using the method of averaging. This is a generalization
of the simple procedure that is commonly 
employed in Hamiltonian dynamics: The Hamiltonian is averaged over certain
``fast'' variables and the resulting averaged Hamiltonian is
used to derive new dynamical equations that presumably describe the ``slow''
motion in a certain averaged sense. The general method is described in
Appendix~\ref{appendixb}, and it is applied to the dissipative 
dynamical system under consideration here in the rest of this paper.

Resonance is a general and significant physical phenomenon and the description of the
state of a physical system while trapped in resonance is of intrinsic importance. The inherent
dynamics at resonance is trivial for a one dimensional oscillator, but is rich in
physical consequences for higher dimensional systems. 
While the general methods
described here could be applied to a wide variety of physical problems, we confine our
attention to a single model.
Our results may, however, 
be of qualitative interest in dealing with the three-body 
problems that arise in the discussions of the origin of the structure 
in the solar system.
 
The present paper relies on the results of our recent work \cite{cmr2}.  We have
repeated here only what is needed for the discussion of the dynamics at
resonance;  for a more precise and complete presentation of the
background material, our papers \cite{cmr,cmr1,cmr2} should be consulted.

Finally, a basic limitation of our model
should be noted. The only damping mechanism that we take into account is
gravitational radiation reaction; this is
consistent with our assumption that the binary consists of Newtonian point
masses moving in vacuum except for the presence of background gravitational
radiation. In this model, a theorem of Neishtadt can be used to show that 
resonance trapping is a rather rare phenomenon~\cite{cmr2}. However, taking due
account of the finite size and structure of astronomical bodies and the existence
of an ambient medium, we would have to include in our model---among other things---the
additional damping effects of tidal friction as well as the various drag effects of
the ambient medium and electromagnetic radiation 
(cf., for instance,~\cite{melita,henrard,beauge,gomes,lai}).
These additional frictional effects could well combine to violate the
condition $N$ of Neishtadt's Theorem~\cite{cmr2,arnold2}. 
Thus, resonance trapping
may not be so rare in astrophysics after all~\cite{bj}. Inclusion of these additional
effects is beyond the scope of our work.
\section{Partial Averaging Near A Resonance}
We will consider the dynamics of the model system~(\ref{MathEQM}) that is
derived in~\cite{cmr2}.
It has the following form when expressed in the Delaunay elements for the
Kepler problem under consideration (cf. Appendix~\ref{appendixa}):
\begin{eqnarray}\label{AAAveraging}
\dot{L} & = & -\epsilon 
\frac{\partial {\cal H}_{\mbox{\scriptsize ext}}}{\partial \ell} +
\epsilon\Delta f_L, \nonumber \\
\dot{G} & = & -\epsilon
\frac{\partial {\cal H}_{\mbox{\scriptsize ext}}}{\partial g} +
\epsilon\Delta f_G \nonumber \\
\dot{\ell}& = &\quad \frac{1}{L^3} 
   + \epsilon \frac{\partial{\cal H}_{\mbox{\scriptsize ext}}}{\partial L}
     +\epsilon\Delta f_{\ell},\nonumber \\ 
\dot{g} & =  & \quad\epsilon 
\frac{\partial {\cal H}_{\mbox{\scriptsize ext}}}{\partial G} +
\epsilon\Delta f_g, 
\end{eqnarray}
where $\epsilon{\cal H}_{\mbox{\scriptsize ext}}$ is the Hamiltonian
corresponding to the external perturbation and
\begin{equation}\label{Hext}
{\cal H}_{\mbox{\scriptsize ext}}=\frac{1}{2}\Omega^2\left[
\alpha {\cal C}(L,G,\ell,g)\cos{\Omega t}+\beta {\cal S}(L,G,\ell,g)\cos(\Omega t+\rho)
\right].
\end{equation}
Here
\begin{eqnarray}\label{CSfs}
{\cal C}(L,G,\ell,g) & = & \frac{5}{2}a^2e^2\cos{2g}
      +a^2\sum_{\nu =1}^\infty (A_\nu\cos{2g}\cos {\nu\ell}
                            -B_\nu\sin{2g}\sin{\nu \ell}),  \nonumber \\
{\cal S}(L,G,\ell,g) & = &\frac{5}{2}a^2e^2\sin{2g}
   +a^2\sum_{\nu =1}^\infty (A_\nu\sin{2g}\cos {\nu\ell}
                            +B_\nu\cos{2g}\sin{\nu \ell}), \nonumber \\
A_\nu & = &\frac{4}{\nu^2e^2}
            \big[2\nu e(1-e^2)J_\nu'(\nu e)-(2-e^2)J_\nu(\nu e)\big], \nonumber \\
B_\nu & = & -\frac{8}{\nu^2e^2}(1-e^2)^{1/2}\,
     \big[e J_\nu'(\nu e)-\nu (1-e^2)J_\nu(\nu e)\big],
\end{eqnarray}
$J_\nu$ is the Bessel function of the first kind of order $\nu$,
and $e=(L^2-G^2)^{1/2}/L$. 
The radiation reaction ``forces" $f_D$, $D\in\{L,G,\ell,g\}$, are certain
complicated functions
of the Delaunay variables given in Appendix~\ref{appendixa}. 
In fact, we will only require the
averages of the $f_D$ given by
\[\bar f_{D}:=\frac{1}{2\pi}\int_0^{2\pi} f_{D}(L, G, \ell, g) \,
 d\ell.\]
These have been computed in~\cite{cmr2}, and are given by 
\begin{eqnarray}\label{fqAve3}
\bar f_{L}&=&-\frac{1}{G^7}\big( 8+\frac{73}{3} e^2+ \frac{37}{12} e^4 \big),
\nonumber \\
\bar f_{G}&=&-\frac{1}{L^3G^4}(8+7e^2),\nonumber \\
\bar f_{\ell}&=&0, \qquad \bar f_{g}=0.
\end{eqnarray}

In order to study the dynamics of the system~(\ref{AAAveraging}) at resonance,
we will apply the method of averaging. We note here that averaging
over the fast angle $\ell$ and the time $t$ gives the correct approximate
dynamics for most initial conditions via Neishtadt's Theorem as explained
in our previous paper~\cite{cmr2}. Here we are interested in the orbits
that are captured into resonance. To study them,  we consider partial
averaging at each resonance.

Let us fix the value of $L$ at the $(m:n)$ resonance, say
$L=L_*$ with $m/L_*^3=n\Omega$, and consider the deviation of $L$ from
the resonance manifold.
To measure this deviation, we introduce a new variable ${\cal D}$ given by
\begin{equation}\label{coordtrans}
 L = L_* + \epsilon^{1/2}\; {\cal D}
\end{equation}
and a new angular variable $\varphi$ by
\[\ell=\frac{1}{{L_*}^3}\: t+\varphi=\frac{n\Omega}{m}t+\varphi.\]
The scale factor $\epsilon^{1/2}$ ensures that, after changing to
the new variables in~(\ref{AAAveraging}), the resulting equations
for $\dot {\cal D}$ and $\dot \varphi$ have the same order in 
the new small parameter $\epsilon^{1/2}$ and therefore the system is in
the correct form for averaging.
These new coordinates are standard choices in the mathematical literature
(for more details see, for example,~\cite{arnold2} or~\cite{wig}).
It is important to emphasize here that the small parameter in the 
actual dynamics is $\epsilon$; however, the small parameter turns out to
be $\epsilon^{1/2}$ in this case for the averaged dynamics.

To effect the coordinate transformation, we use the expansion
\begin{equation}\label{tayL}
\frac{1}{{L}^3}=\frac{1}{{L_*}^3}\Big[1-\epsilon^{1/2}\frac{3{\cal D}}{L_*}
  +\epsilon\frac{6{\cal D}^2}{L_*^2} +O(\epsilon^{3/2})\Big]
\end{equation}
and find
\begin{eqnarray}\label{AA1Averaging}
\dot{{\cal D}} & = & -\epsilon^{1/2} F_{11}-\epsilon {\cal D} F_{12}+O(\epsilon^{3/2}),
 \nonumber \\
\dot{G} & = & -\epsilon F_{22}+O(\epsilon^{3/2}), \nonumber \\
\dot{\varphi} & = & -\epsilon^{1/2}\frac{3}{L_*^4}{\cal D}
     + \epsilon \big( \frac{6}{L_*^5}{\cal D}^2+F_{32}\big)+O(\epsilon^{3/2}),\nonumber \\
\dot{g} & = &  \quad\epsilon F_{42}+O(\epsilon^{3/2}), 
\end{eqnarray}
where the
$F_{ij}(G, n\Omega t/m+\varphi,g,t)$ 
are defined such that the first index refers to the equation in which
it appears and the second index refers to the perturbation order in powers
of $\epsilon^{1/2}$ that is employed. 
These quantities are given by
\begin{eqnarray}\label{Fij}
F_{11}&:=&\frac{\partial{\cal H}_{\mbox{\scriptsize ext}}}{\partial
 \ell}(L_*,G,\frac{n\Omega}{m}t+\varphi,g,t)
             -\Delta f_L(L_*,G,\frac{n\Omega}{m}t+\varphi,g), \nonumber \\
F_{12}&:=&\frac{\partial^2{\cal H}_{\mbox{\scriptsize ext}}}{\partial L\partial \ell}
                             (L_*,G,\frac{n\Omega}{m}t+\varphi,g,t)
         -\Delta \frac{\partial f_L}{\partial
 L}(L_*,G,\frac{n\Omega}{m}t+\varphi,g),\nonumber \\
F_{22}&:=&\frac{\partial{\cal H}_{\mbox{\scriptsize ext}}}{\partial
 g}(L_*,G,\frac{n\Omega}{m}t+\varphi,g,t)
             -\Delta f_G(L_*,G,\frac{n\Omega}{m}t+\varphi,g),\nonumber \\
F_{32}&:=&\frac{\partial{\cal H}_{\mbox{\scriptsize ext}}}
   {\partial L}(L_*,G,\frac{n\Omega}{m}t+\varphi,g,t)
             +\Delta f_\ell(L_*,G,\frac{n\Omega}{m}t+\varphi,g), \nonumber\\
F_{42}&:=&\frac{\partial{\cal H}_{\mbox{\scriptsize ext}}}{\partial
 G}(L_*,G,\frac{n\Omega}{m}t+\varphi,g,t)
             +\Delta f_g(L_*,G,\frac{n\Omega}{m}t+\varphi,g).
\end{eqnarray}

The system~(\ref{AA1Averaging}) is $2\pi m/\Omega$ periodic in
the temporal variable---since $\cal H_{{\mbox{\scriptsize ext}}}$ is
$2\pi /\Omega$ periodic in time---and is in time-periodic standard form.
Anticipating our intention to average to second order, we will apply an
averaging transformation (for a detailed exposition see~\cite{wig}).
It is the characteristic property of this transformation that it automatically
renders system~(\ref{AA1Averaging}) in a form such that to lowest order
the new system is exactly the first order averaged system and the second order
averaged system can be simply obtained by averaging the new 
system (cf. Appendix~\ref{appendixb}).
To obtain the desired transformation, we define
\[
\bar{F}_{ij}:= \frac{\Omega}{2\pi m}
     \int_0^{2\pi m/\Omega} F_{ij}(G,\frac{n\Omega}{m}s+\varphi,g,s)\,ds,
\]
and the deviation from the mean for $F_{11}$
\begin{equation}\label{avertran}
\lambda(G,\varphi,g,t):=F_{11}(G,\frac{n\Omega}{m}t+\varphi,g,t)-\bar{F}_{11}.
\end{equation}
Furthermore, we define $\Lambda(G,\varphi,g,t)$ to be the antiderivative of
$\lambda(G,\varphi,g,t)$ with respect to $t$ with the additional property
that the average of $\Lambda$ should vanish, i.e.
\[
\int_0^{2\pi m/\Omega} \Lambda(G,\varphi,g,s)\,ds=0.
\]
Moreover, we note that
both  $\lambda$ and $\partial\Lambda/\partial\varphi$ have
zero averages.  Our averaging transformation is given by
\[
{\cal D}=\widehat {{\cal D}}
 -\epsilon^{1/2}\Lambda(\widehat{G},\widehat{\varphi},\widehat{g},t),\quad
G=\widehat{G},\quad \varphi=\widehat{\varphi},\quad g=\widehat{g}.
\]
The averaging transformation is constructed so that its average becomes
the identity transformation.

Let us observe that if $G$, $\varphi$, and $g$ depend on $t$ as solutions of the
system~(\ref{AA1Averaging}), then
\[
\dot\Lambda=\lambda-\epsilon^{1/2}\Big(\frac{3{\cal D}}{L_*^4}\Big)
     \frac{\partial\Lambda}{\partial\varphi} +O(\epsilon).
\]
After applying the averaging transformation, we find that the
system~(\ref{AA1Averaging}) takes the form
\begin{eqnarray}\label{AA2Averaging}
\dot{\widehat{{\cal D}}} & = & -\epsilon^{1/2} \bar{F}_{11}
  -\epsilon \widehat{{\cal D}} \Big( F_{12}
     +\frac{3}{L_*^4}\frac{\partial\Lambda}{\partial\varphi}\Big)
      +O(\epsilon^{3/2}), \nonumber \\
\dot{\widehat{G}} & = & -\epsilon F_{22}+O(\epsilon^{3/2}), \nonumber \\
\dot{\widehat{\varphi}} & = & -\epsilon^{1/2}\frac{3}{L_*^4}\widehat{{\cal D}}
     + \epsilon \Big(\frac{6}{L_*^5}{\widehat{{\cal D}}}^2+F_{32}
       +\frac{3}{L_*^4}\Lambda\Big)+O(\epsilon^{3/2}), \nonumber \\
\dot{\widehat{g}} & = &\quad \epsilon F_{42}+O(\epsilon^{3/2}).
\end{eqnarray}
Finally, we drop the $O(\epsilon^{3/2})$ terms in~(\ref{AA2Averaging}) and 
average the remaining truncated system 
to obtain the {\em second order partially averaged system}
\begin{eqnarray}\label{2ndoave}
\dot{\widetilde{{\cal D}}}
   & = & -\epsilon^{1/2} \bar{F}_{11}
    -\epsilon\widetilde{{\cal D}}\bar{F}_{12},\nonumber \\
\dot{\widetilde{G}} & = & -\epsilon \bar{F}_{22}, \nonumber \\
\dot{\widetilde{\varphi}} & = & -\epsilon^{1/2}\frac{3}{L_*^4}\widetilde{{\cal D}}
     + \epsilon \big(\frac{6}{L_*^5}{\widetilde{{\cal D}}}^2+\bar{F}_{32}\big),\nonumber \\
\dot{\widetilde{g}} & = &  \quad \epsilon \bar{F}_{42}.
\end{eqnarray}
This system is the averaged form of system~(\ref{AA1Averaging}) 
after dropping its $O(\epsilon^{3/2})$ terms;
however, this coincidence is fortuitous in this case.
In general, one has to employ an averaging
transformation in order to obtain the second order averaged system.

To explain the evolutionary dynamics at resonance, we will
replace the actual dynamical
equations by the second order partially averaged system. As explained in 
Appendix~\ref{appendixb}, this is a reasonable approximation over a limited time-scale.
We remark that although the second order partially averaged system will be used
to explain some of the features of  our model system near its resonances,
the actual dynamics predicted by our model is certainly much more complex than the
averaged equations reveal. In particular, we expect that near the resonances---perhaps
in other regions of the phase space as well---there are chaotic invariant sets and, 
therefore, transient chaotic motions~\cite{cmr2}. 
The averaged system is nonlinear and could in general exhibit chaos; however,
we have not encountered chaos in the second order averaged equations obtained 
from the model under consideration here. 

It remains to compute $\bar{F}_{ij}$, where $F_{ij}$ are defined
in~(\ref{Fij}). For this, we  recall equation~(\ref{Hext})
and set $\ell=n\Omega t/m+\widetilde{\varphi}$,
expand the trigonometric terms in the Fourier series using trigonometric
identities, and then average over the variable $t$. The required 
averages involving the external perturbation can be computed from the average of
${\cal H}_{\mbox{\scriptsize ext}}$. In fact, after some computation, we find that
\begin{eqnarray*}
\bar{{\cal H}}_{\mbox{\scriptsize ext}}&:=&\frac{\Omega}{2\pi m}\int_0^{2\pi m/\Omega}
  {\cal H}_{\mbox{\scriptsize ext}}(L_*,\widetilde{G},
   \frac{n\Omega}{m}t+\widetilde{\varphi},\widetilde{g},t)\,dt\\
&=&
T_c(L_*,\widetilde{G},\widetilde{g})\cos{m\widetilde{\varphi}}
+T_s(L_*,\widetilde{G},\widetilde{g})\sin{m\widetilde{\varphi}},
\end{eqnarray*}
where, for $n=1$,
\begin{eqnarray}\label{tcts}
T_c & := & \frac{L^4\Omega^2}{4}\big[
\alpha A_m(e)\cos{2g}+\beta A_m(e)\sin{2g}\cos{\rho}
                         -\beta B_m(e)\cos{2g}\sin{\rho}\big],\nonumber \\
T_s & := & \frac{L^4\Omega^2}{4}\big[-\alpha B_m(e)\sin{2g}+\beta
 A_m(e)\sin{2g}\sin{\rho}
                         +\beta B_m(e)\cos{2g}\cos{\rho}\big],
\end{eqnarray}
and for $n\ne 1$, $\bar{{\cal H}}_{\mbox{\scriptsize ext}}=0$ so
that in this case we can define  $T_c=T_s=0$.

The averages of $f_D$, $D\in\{L,G,\ell,g\}$,  are given in~(\ref{fqAve3}).
The terms involving radiation damping
 that appear in the partially averaged system are obtained from
these expressions by Taylor expansion about the resonant orbit using
equation~(\ref{coordtrans}). For example, we will use
\[
\Gamma(G):=\left.\bar{f}_L\right|_{L=L_*}=
-\left.\frac{1}{G^7}\left(8 + \frac{73}{3} e^2 + \frac{37}{12} e^4 \right)
\right|_{e^2=1-G^2/L_*^2}
\]
and the average of $\partial f_L/\partial L$ at resonance given by
\[
\left.\frac{\partial}{\partial L} \bar{f}_L \right|_{L=L_*}=
\left.-\frac{1}{3 L_*^3 G^5}(146+37 e^2)\right|_{e^2=1-G^2/L_*^2}.
\]
The second order partially averaged system~(\ref{2ndoave}) is
thus given explicitly by
\begin{eqnarray}\label{ex2ndoave}
\dot {\cal D} &=& -\epsilon^{1/2}\Big[-mT_c\sin{m\varphi}+mT_s\cos{m\varphi}
  +\frac{\Delta}{G^7}\Big(8 + \frac{73}{3} e^2 + \frac{37}{12} e^4 \Big)\Big]
\nonumber \\
&&\quad-\epsilon {\cal D} \Big[-m\frac{\partial T_c}{\partial L}\sin{m\varphi}
                              +m\frac{\partial T_s}{\partial L}\cos{m\varphi}
  +\frac{\Delta}{3L_*^3 G^5}\Big(146 + 37 e^2 \Big)\Big], \nonumber \\
\dot G &=& -\epsilon\Big[\frac{\partial T_c}{\partial g}\cos{m\varphi}
                              +\frac{\partial T_s}{\partial g}\sin{m\varphi}
  +\frac{\Delta}{L_*^3 G^4}(8 + 7 e^2 )\Big], \nonumber \\
\dot \varphi &=& -\epsilon^{1/2}\frac{3}{L_*^4} {\cal D}+\epsilon\Big(
  \frac{6}{L_*^5}{\cal D}^2+\frac{\partial T_c}{\partial L}\cos{m\varphi}
   +\frac{\partial T_s}{\partial L}\sin{m\varphi}\Big), \nonumber \\
\dot g &=& \:\epsilon\Big(\frac{\partial T_c}{\partial G}\cos{m\varphi}
              +\frac{\partial T_s}{\partial G}\sin{m\varphi}\Big),
\end{eqnarray}
where we have dropped the tildes.
It is clear that $L$ in the
expressions involving $T_c$, $T_s$, and $e$ must be replaced by $L_*$, the value of
$L$ at resonance.

Having derived the equations for the averaged dynamics~(\ref{ex2ndoave}),
we now turn our attention to the consequences of these equations and the comparison
of predictions based on them with the actual dynamics given by~(\ref{MathEQM}). 

\section{First Order Averaged Dynamics}\label{drt}
The first order partially averaged system, obtained from~(\ref{ex2ndoave}) 
by dropping the
$O(\epsilon)$ terms, is given by
\begin{eqnarray}\label{foav}
\dot{\cal D}&=&-\epsilon^{1/2}\Big[-mT_c\sin{m\varphi}+mT_s\cos{m\varphi}
       -\Delta \Gamma(G)\Big],\nonumber\\
\dot G&=&0,\nonumber\\
\dot \varphi&=&-\epsilon^{1/2}\Big(\frac{3}{L_*^4}\Big) {\cal D}, \nonumber\\
\dot g&=&0.
\end{eqnarray}
In this approximation, the variables $G$ and $g$ are constants fixed
by the initial conditions while the remaining system in ${\cal D}$ and $\varphi$
is equivalent to a pendulum-type equation with torque; namely,
\begin{equation}\label{pendulum}
\ddot\varphi+\frac{3\epsilon}{{L_*}^4}(mT_c\sin{m\varphi}-mT_s\cos{m\varphi})
   =-\frac{3\epsilon}{{L_ *}^4}\: \Delta\: \Gamma(G).
\end{equation}
We also have a second order differential equation---a harmonic
oscillator with slowly varying frequency---for the deviation
${\cal D}$ given by
\begin{equation}\label{fodeveq}
\ddot {\cal D}+\epsilon w^2 {\cal D}=0,
\end{equation}
where
\begin{equation}\label{w}
w:=\Big[\frac{3m^2}{L_*^4}(T_c\cos{m\varphi}+T_s\sin{m\varphi})\Big]^{1/2}.
\end{equation}
These results can be formally justified if they are recast in terms of a 
new temporal variable given be $\epsilon^{1/2}t$; however, we use $t$ in
order to facilitate comparison with the actual dynamics.

To show that~(\ref{fodeveq}) is an oscillator, we must show that $w$ is a
real number.
To this end, we suppose that during capture into the resonance the orbit
is near an elliptic region of the first order partially averaged system~(\ref{foav}).
This system is Hamiltonian with an effective energy of the form
\[ 
-\epsilon^{1/2}\Big[\frac{1}{2}\Big(\frac{3}{L_*^4}\Big) {\cal D}^2+U(\varphi)\Big],
\]
where $U$ represents  the effective potential energy  given by
\[
U:=-(T_c\cos{m\varphi}+T_s \sin{m\varphi})+(\Delta\Gamma)\varphi.
\]
If $({\cal D}, \varphi)=(0,\varphi_0)$ is an elliptic rest point of the 
first order partially averaged system; then,
$U'(\varphi_0)=0$ and $U''(\varphi_0)>0$, where a prime denotes 
differentiation with respect to $\varphi$.  It follows that
\[U''(\varphi_0)= m^2(T_c\cos{m\varphi_0}+T_s\sin{m\varphi_0}),\]
so that $w_0$, $w_0:=[3 U''(\varphi_0)]^{1/2}/L_*^2$, must be real.

To show that the frequency $\xi:=\epsilon^{1/2}w$ is slowly varying,
we differentiate this frequency with respect to time to
obtain
\[
\dot\xi=-\epsilon\frac{9m^3}{2L_*^8w}(T_s\cos{m\varphi}-T_c\sin{m\varphi}){\cal D}+
             O(\epsilon^{3/2}).
\]
It follows from the first order averaged equations that
$\varphi-\varphi_0$ is expected to be of order unity; thus,
$\xi$ is nearly constant over a time-scale of order
$\epsilon^{-1/2}$ since $\dot\xi$ is proportional to $\epsilon$.
In particular, ${\cal D}$ varies on the time-scale $\epsilon^{1/2}t$.
Inspection of equations~(\ref{pendulum}) and~(\ref{fodeveq}) 
reveals that while ${\cal D}$
is predominantly a simple harmonic oscillator 
with frequency $\xi_0=\epsilon^{1/2}w_0$, 
the motion of $\varphi$ in time could be
rather complicated involving essentially all harmonics
of $\xi_0$. Therefore, $\varphi$ cannot in general
be considered a slow variable in time. 

Librational motions (periodic motions in the phase plane) of the pendulum-type
averaged equation~(\ref{pendulum}) correspond to orbits of the original system
that are captured into the resonance. On the other hand, if there are no
librational motions, then all orbits pass through the resonance.
Thus, we observe a necessary condition for
capture: the pendulum system must have rest points in its phase plane.
The rest points of the first order averaged system are given by
${\cal D}=0$ and $\varphi=\varphi_0$, with
\begin{equation}\label{resteq}
R\sin(m\varphi_0+\eta)=-\Delta \Gamma,
\end{equation}
where $mT_c=R\cos{\eta}$ and $mT_s=-R\sin{\eta}$.
As an immediate consequence, we have the following proposition:
{\em For the $(m:n)$ resonance, if $n\ne 1$, there is no capture into resonance}.
On the other hand,  for $n=1$, 
if there are librational motions, then
we must have $\Delta|\Gamma/R| \le 1$.
These observations suggest that in the presence of sufficiently strong
radiation damping, i.e.  for $\Delta$ sufficiently large, there are no
librational
motions and, as a result, each orbit will pass through all the resonant
manifolds that it encounters. 
In particular, since $L=a^{1/2}$,  the semimajor
axis of the corresponding osculating ellipse will collapse to zero.
Moreover, capture is only possible for $(m:n)$ resonances with $n=1$ when
$\Delta|\Gamma/R| \le 1$. In this case, if $\Delta|\Gamma/R| < 1$, then
the rest points in the phase plane of the pendulum-type equation~(\ref{pendulum})
at the resonance are all nondegenerate. This is precisely Neishtadt's
condition $B$.

The quantities $\Gamma$ and $R$ 
depend (nonlinearly) on the variables $(L, G, g)$ as well as
the parameters of the system; therefore, the precise range of their 
ratio $\Gamma/R$ is difficult to
specify in general.  However, the value of this ratio
can be determined numerically. In fact, to find orbits that are captured
into resonance as displayed in Figures~1--3, we use the first order averaged
system. A rest point of the first order system corresponds
to a ``fixed'' resonant orbit with ${\cal D}=0$ and constant
$G$, $g$, and $\varphi=\varphi_0$. The main characteristics of these orbits
do not change with respect to the slow variable $\epsilon^{1/2}t$;
that is, the resonant orbits are essentially fixed only over a time-scale of
order $\epsilon^{-1/2}$. 
After choosing $G$ and $g$, 
we solve for $\varphi_0$ at a rest point, 
and then test to see if the resulting resonant orbit corresponds
to a libration point of the first order averaged system (e.g. $w_0^2$
must be positive). If it does, we 
convert the point with coordinates
$(L_*, G,\varphi_0,g)$ to polar coordinates and then numerically integrate
our model~(\ref{MathEQM}) backward in time from this starting value. 
When the backward integration in time is carried
out over a sufficiently long time interval, the orbit is expected to
leave the vicinity of the resonance.  
After this occurs, we integrate forward 
in time to obtain an orbit of~(\ref{MathEQM}) that is captured
into the resonance.

The first order partially averaged system~(\ref{foav}) is
Hamiltonian; therefore,
we do not expect that its dynamics would  give a complete picture of the average
dynamics near a resonance for our {\em dissipative} system. In particular, the
librational motions of the pendulum-like system  are not structurally stable.
Indeed, the phase space for the full four dimensional system~(\ref{foav}) 
is foliated by invariant two dimensional subspaces parametrized by the 
variables $G$ and $g$.
If a librational motion exists in a leaf of the foliation, then
the corresponding rest points,  the elliptic rest point and the
hyperbolic rest point (or points) at the boundary of the librational region for the
associated pendulum-like system,  are degenerate in the four
dimensional phase space of the first order averaged system; 
their linearizations have
two zero eigenvalues corresponding to the directions normal to the leaf.
In the second order partially averaged system, viewed as a small perturbation
 of~(\ref{foav}),
we expect that only a finite number of the degenerate rest points survive. These
correspond to the continuable periodic solutions described in~\cite{cmr2}.
We expect the corresponding perturbed rest points
to be hyperbolic. As a result, they persist even
when higher order effects are added. The dimensions and positions of the
stable and unstable manifolds of these rest points determine the average
motion near the resonance. 
In particular, if one of these rest points is a hyperbolic sink,
then there is an open set of initial conditions that correspond to orbits
{\em permanently} captured into resonance.
Our numerical experiments have provided no evidence for this behavior.
However, another possibility---that is consistent with our 
numerical experiments---is that a rest point of the perturbed system
in phase space is stable
along two directions but unstable in the remaining two directions.
Thus a trajectory with initial point near the corresponding stable manifold will
be captured into resonance and undergo librational motions near the resonant
manifold until it spirals outward along the unstable manifold.
\section{Dynamical Evolution During Resonance}\label{dedr}
The dynamics of system~(\ref{ex2ndoave}) is expected to 
give a close approximation of the
near resonance behavior of our original model~(\ref{MathEQM}) over
a time-scale of order $\epsilon^{-1/2}$. In particular,
a basic open problem is the following: Determine the positions of the rest points 
and the corresponding stable and unstable manifolds of~(\ref{ex2ndoave}).
A solution of this problem would provide the information needed to analyze 
the most important features of the behavior of the orbits that pass through the 
resonance as well as those that are captured by the resonance. Unfortunately, 
rigorous analysis of the dynamics in 
the four dimensional phase space of system~(\ref{ex2ndoave}) 
seems at present to be very difficult. Thus, in lieu of a complete and rigorous
analysis of the dynamics,  we will show how to obtain useful information from 
approximations of the second order partially averaged system. 

A typical example of the behavior of the  orbit as it passes through a
$(1:1)$ resonance is depicted in Fig.~\ref{fig1}. The semimajor axis undergoes
librations of increasing amplitude around the resonant value. Furthermore, 
the eccentricity of the orbit generally decreases while the orbit is trapped.

We propose to analyze the oscillations of ${\cal D}$ by using the
$O(\epsilon)$ approximation of the second order differential equation for ${\cal D}$ 
derived from the second order partially averaged equations~(\ref{ex2ndoave}).
In fact,  a simple computation 
yields the following differential equation
\begin{equation}\label{devos}
\ddot{\cal D}+\epsilon \gamma \dot{\cal D}+\epsilon w^2 {\cal D}=0,
\end{equation}
where
\begin{equation}\label{gam25}
\gamma:=m\Big(
  \frac{\partial T_s}{\partial L}\cos{m\varphi}
    -\frac{\partial T_c}{\partial L}\sin{m\varphi}\Big)
      +\frac{\Delta}{3L^3G^5}(146+37e^2)
\end{equation}
is evaluated at $L=L_*$ in equation~(\ref{devos})
and $w$ is given by equation~(\ref{w}).
Similarly, the second order partially averaged system~(\ref{ex2ndoave})
can be used to derive an equation for the temporal evolution of $\varphi$
to order $\epsilon$; however, the resulting equation turns out to
be identical with equation~(\ref{pendulum}).

It is clear from inspection of the differential equation~(\ref{devos}) that
$L=L_*+\epsilon^{1/2}{\cal D}$ oscillates about its resonant
value $L_*$ with a libration frequency given approximately by 
$\xi_0=\epsilon^{1/2} w_0$. 
The magnitude of this frequency is in agreement with our numerical results to
within a few percent for the case of resonance trapping depicted in Fig.~\ref{fig1}.
Moreover,
the amplitude of the oscillations will increase (decrease) if $\gamma<0$
($\gamma>0$). 

The sign of $\gamma$ varies with the choice of parameters, variables, and
the order of the resonance. Our numerical experiments---which have 
by no means been exhaustive---indicate that the $(1:1)$ resonance for the
linearly polarized incident wave with $\alpha=1$ and $\beta=0$
is special in that $\gamma$ is negative at the 
elliptic rest points of the first order averaged system. 
However, for a resonance with order $m>1$, $\gamma$ is
not of fixed sign. Nevertheless, our numerical experiments indicate that
the amplitude of librations generally increases over a long time-scale 
(cf.  Fig.~\ref{fig2}).
Figures~\ref{fig2} and~\ref{fig3} illustrate the apparent richness of the
evolutionary dynamics near resonances with $m>1$.
It should be emphasized that the damped (or anti-damped)
oscillator~(\ref{devos}) approximates the
actual dynamics only over a time-scale of order $\epsilon^{-1/2}$ 
(cf. Appendix~\ref{appendixb}).

To obtain an approximate equation for the envelope of the oscillations,
we consider a time-dependent change of variables for the oscillator~(\ref{devos})
defined by the relation
\begin{equation}\label{Deq} 
{\cal D}=Ve^{-(\epsilon/2)\int_0^t\gamma(s)\,ds}.
\end{equation}
After a routine calculation, we obtain the differential equation
\begin{equation}\label{harmos}
\ddot V+\epsilon w^2 V=O(\epsilon^{3/2}).
\end{equation}
Thus, to order $\epsilon$,  $V(t)$ in equation~(\ref{Deq}) is 
the solution of a harmonic oscillator equation with a 
slowly varying frequency $\xi=\epsilon^{1/2}w$. In fact,
equation~(\ref{harmos}) is identical to the oscillator~(\ref{fodeveq}) 
with frequency $\xi$ obtained from the first order
averaged system.  
The solution ${\cal D}(t)$ of the second order equation~(\ref{devos})
in this approximation
is obtained by modulating the amplitude of $V$.

We note that if $\gamma$ and $w$ are constants, then formula~(\ref{Deq})
reduces to 
\[
{\cal D}(t)={\cal A}e^{-\epsilon\gamma t/2} \cos(\epsilon^{1/2}wt+\tau),
\]
where ${\cal A}$ and $\tau$ are constants depending on the initial conditions;
this is the standard result for the damped 
(or anti-damped) harmonic oscillator with constant coefficients.

Equation~(\ref{Deq}) certainly agrees qualitatively with the numerical
experiments reported in Figures~\ref{fig1}--\ref{fig3}. 
To check the accuracy quantitatively, we numerically integrated 
system~(\ref{MathEQM}) and simultaneously evaluated
the integral of $\gamma$ in order
to obtain an approximation for the envelope
of ${\cal D} (t)$. 
Over an interval of time of length $691$, which is of order  $\epsilon^{-1/2}$, 
the value of ${\cal D}$ at its maximum predicted from
equation~(\ref{Deq}) differs from the value obtained by numerical integration
of the differential equations by only a few percent (actually $2.5\%$). 
The error in the comparison of the two values for the envelope of  ${\cal D}$ 
grows slowly over time; in fact, it remains less than $20\%$ over a time interval 
of length $9000$.

The evolutionary dynamics near resonance
is characterized by librational motions described
above as well as variations in angular momentum.
Our numerical experiments suggest considerable variation in the
angular momentum while the orbit is trapped in resonance.
The remainder of this section is devoted to a theoretical explanation
of this phenomenon.

During the time that an orbit is captured into resonance, 
we expect that on average it will reside near an elliptic rest point of the
first order partially averaged system.
Thus, we expect the dynamical variables to satisfy approximately
equation~(\ref{resteq}) while the orbit is trapped. 
We will show how to determine the dynamical behavior of the angular
momentum $G$ under this assumption. 
In particular, for the $(1:1)$ resonance, we claim that $\dot G>0$; that is, 
the orbital angular momentum increases. 

It is easy to show that
\begin{eqnarray*}
\frac{\partial T_c}{\partial g}-2 T_s&=&
  -\frac{1}{2L_*^2}[\alpha \sin 2g-\beta\cos(2g + \rho)]S_m(e), \\
\frac{\partial T_s}{\partial g}+2 T_c&=&
\frac{1}{2L_*^2}[\alpha \cos 2g+\beta\sin(2g + \rho)]S_m(e),
\end{eqnarray*}
where 
\[ S_m(e):=m^2[A_m(e)-B_m(e)].\]
After substituting these identities as well as the condition~(\ref{resteq})
into the expression for 
$\dot G$ in~(\ref{ex2ndoave}),
we find that 
\begin{equation}\label{gdot}
\dot G= -\epsilon [P_m(e)+L_*^{-2} S_m(e) Q_m],
\end{equation}
where, after some manipulation using the relation  $e=(1-G^2/L^2_*)^{1/2}$,
\begin{eqnarray*}
P_m(e)&:=&\frac{\Delta}{G^7}\Big[(1-e^2)^{3/2}(8+7 e^2)
  -\frac{2}{m}\big( 8+\frac{73}{3} e^2+ \frac{37}{12} e^4 \big)\Big],\\
Q_m&:=&\frac{1}{2} [\alpha\sin(m\varphi-2g)+\beta\cos(m\varphi-2g-\rho)].
\end{eqnarray*}
The function $S_m(e)\to 0$ as $e\to 0$; therefore, 
it is clear that $\dot G>0$ near $e=0$ as long as $P_m(e)<0$. 
We have $P_1(e)<-8\Delta/G^7$ for $0<e<1$. For $m=2$, 
we have $P_2(e)<0$, but
$P_2(0)=0$, while for $m>2$, we have $P_m(e)>0$ near $e=0$.
Our conclusion is that near the $(1:1)$ resonance, we can expect
$\dot G>0$ during capture provided the osculating
ellipse is not too eccentric. The other cases are more delicately balanced.
However, for a specific choice of parameters and for a specific
resonance, one can use equation~(\ref{gdot}) to predict
the behavior of $G$ near the resonance.

Consider, for instance, the system~(\ref{MathEQM}) with the following parameter values:
$\epsilon=10^{-4}$, $\Delta=\delta/\epsilon=10^{-3}$,
$\alpha=1$, $\beta=0$, $\rho=0$, and $\Omega= 1$. 
The orbit, as depicted in Figure~\ref{fig1}, with
initial conditions given by 
\[
(p_r,p_\theta,r,\theta)=(0.2817, 0.6389, 1.6628, 2.9272)
\]
appears to be trapped in $(1:1)$ resonance with a sojourn time of approximately
$40000$.
Moreover,
during this period of time 
the angular momentum appears to increase from approximately
$0.66$ to $0.95$.
To demonstrate that our theoretical scheme using the second order partially
averaged system agrees with our numerical results, let us determine the 
time-scale over which the angular momentum changes from $0.78$ to $0.95$.
According to the numerical integration of system~(\ref{MathEQM}) this occurs
over about $32700$ time units, whereas our approximate 
analysis of the second order partially
averaged system predicts a time-scale of
\[
-\frac{1}{\epsilon}\int_{0.78}^{0.95} 
\big[P_1\big((1-G^2/L^2_*)^{1/2}\big)\big]^{-1}\,dG\approx 34200,
\] 
a time that is within less than $5\%$ of the value given by our numerical experiment. 
In this integral, we have used the fact that in equation~(\ref{gdot}),
the term $L_*^{-2}S_1(e)Q_1$ is small compared with $P_1(e)$. 

It should be clear from the results of this section that---by employing
the second order partially averaged system---we have been able
to provide theoretical explanations for the behaviors of semimajor axis
$a$ and eccentricity $e$ of the orbit that is trapped in resonance.
The basic dynamical equations for our model are only valid to order $\epsilon$,
since physical effects of higher order have been neglected in equation~(\ref{MathEQM}); 
therefore,
it would be inappropriate to go beyond the second order averaged system in
this case. The averaging method is general, however, and can be carried
through to any order.
\section{Conclusion}\label{c}
Some of the observed structures in the solar system are the direct results of
evolutionary dynamics while trapped in resonance
(see \cite{melita}, \cite{beauge}, and \cite{gomes}).
We have argued in this paper that such dynamics can in general be explained using
the method of averaging. In particular, the main aspects of the
evolution of the dynamical system during the time that it is locked in resonance
can be understood on the basis of the second order partially averaged dynamics.
We have illustrated these ideas using a particular model involving a Keplerian
binary system that emits and absorbs gravitational radiation.
\appendix
\section{Delaunay equations of motion}\label{appendixa}
The Delaunay equations of motion for our system are given in~\cite{cmr2} as
\begin{eqnarray}\label{D2EQM}
\frac{dL}{dt} & = & -\epsilon \left(\frac{\partial{\cal C}}{\partial \ell}
\phi(t) +
\frac{\partial{\cal S}}{\partial \ell} \psi(t)\right) +  \delta f_L,
\nonumber \\
\frac{dG}{dt} & = & -\epsilon \left(\frac{\partial{\cal C}}{\partial g}
\phi(t) +
\frac{\partial{\cal S}}{\partial g} \psi(t)\right) +  \delta f_G,
\nonumber \\
\frac{d\ell}{dt} & = & \omega + \epsilon \left(\frac{\partial{\cal C}}{\partial
L} \phi(t) +
\frac{\partial{\cal S}}{\partial L} \psi(t)\right)
+  \delta f_{\ell}, \nonumber \\
\frac{dg}{dt} & = & \epsilon \left(\frac{\partial{\cal C}}{\partial G}
\phi(t) +
\frac{\partial{\cal S}}{\partial G}\psi(t)\right) +  \delta f_g,
\end{eqnarray}
where
\[
\phi(t):=\frac{1}{2} \alpha\Omega^2\cos\Omega t,\qquad
\psi(t):=\frac{1}{2} \beta\Omega^2\cos(\Omega t+\rho),
\]
${\cal C}$ and ${\cal S}$ are given by equation~(\ref{CSfs}) and 
\begin{eqnarray}\label{fequations}
 f_L & = &
\frac{4}{Lr^3}\left[1-\frac{16}{3}\frac{L^2}{r} +
\left(\frac{20}{3}L^2- \frac{17}{2}G^2\right)\frac{L^2}{r^2}
+\frac{50}{3}\frac{L^4G^2}{r^3}-\frac{25}{2}\frac{L^4G^4}{r^4}\right],
\nonumber \\
f_G & = & -\frac{18 G}{L^2r^3}\left(1 -\frac{20}{9}\frac{L^2}{r}
+\frac{5}{3}\frac{L^2G^2}{r^2}\right), \nonumber \\
f_{\ell} & = & \frac{2\sin{\hat{v}}}{eL^3Gr^2}\left[4e^2+
\frac{1}{3}
\left(73G^2-40L^2\right)\frac{1}{r}
-2\left(1+\frac{70}{3}L^2-\frac{29}{2}G^2\right)\frac{G^2}{r^2}
\right. \nonumber \\
&& \hspace*{.5in}\left.-\frac{25}{3}\frac{L^2G^4}{r^3}
+25\frac{L^2G^6}{r^4} \right], \nonumber \\
f_g & = &
-\frac{2\sin{\hat{v}}}{eL^2r^3}\left[11+
\left(7G^2-\frac{80}{3}L^2\right)\frac{1}{r}-
\frac{25}{3}\frac{L^2G^2}{r^2}+25\frac{L^2G^4}{r^3}\right].
\end{eqnarray}
The $f_D$ can be expressed, in principle, in terms
of Delaunay elements.  In fact, expressions of the form $r^{-q}$ and 
$r^{-q}\sin{\widehat{v}}$ can be expanded as Fourier series in terms of the mean
anomaly $\ell$ (see \cite{Kov}, p. 54).  For example, the following results are known
\cite{Kov,watson}
\begin{eqnarray}\label{watexpan}
\frac{L^2}{r} & = &  1+2 \sum_{n=1}^{\infty}J_n(ne) \cos{n\ell},  \\
\frac{L^4}{r^2}\sin{\hat{v}} & = & 
 \frac{(1-e^2)^{1/2}}{e} \sum_{n=1}^{\infty}2n  J_n(ne) \sin{n\ell}.
\end{eqnarray}
For the calculation of the other similar 
expressions---and eventually the $f_D$---see the methods of 
classical celestial mechanics involving the Bessel
functions described in~\cite{Kov} and~\cite{watson}.
Only the {\em averages} of the $f_D$---calculated in~\cite{cmr2} and
given in equation~(\ref{fqAve3})---are
needed for the purposes of this paper.
\section{the method of averaging}\label{appendixb}
In this section, we give a brief introduction to the method of averaging as it is
used in this paper. More detailed accounts can be found in many excellent sources.
We suggest references \cite{arnold2},\cite{murdock}, \cite{chirikov},
\cite{wig},  and~\cite{sv},
as they offer a range of mathematical sophistication
as well as diverse points of view. 

The method of averaging is usually applied to differential equations in one
of several standard forms.
In celestial mechanics, perhaps the most common standard form
is ``angular standard form'' 
\[ \dot I=\epsilon f(I,\theta),\qquad \dot \theta=\omega(I)+\epsilon g(I,\theta),\]
where $I$ is an $m$-dimensional variable, $\theta$ is an $n$-dimensional
angular variable, both $f$ and $g$ are $2\pi$ periodic in $\theta$, and
$\epsilon$ is a small parameter.
This form is often obtained after a model of disturbed two-body motion is expressed
in the action-angle variables appropriate for the
unperturbed Kepler problem as in equation~(\ref{AAAveraging}). 
The second form is ``time periodic standard form''
\begin{equation}\label{tpsf}
 \dot x=\epsilon F(x,t),
\end{equation}
where $x$ is an $m$-dimensional variable and $F$ is $T$-periodic in time.
This form is obtained for many different physical problems. 
In particular, in our case --- i.e. equation~(\ref{AA1Averaging})---
the small parameter is $\epsilon^{1/2}$ and the equation of motion is a Taylor
series in $\epsilon^{1/2}$.
Of course, the time periodic standard form can be viewed as a special
case of the angular standard form by introducing a new one dimensional angular variable:
\[
\dot x=\epsilon F^*(x,\theta),\qquad \dot\theta=\frac{2\pi}{T}. 
\]
However, it is useful to separate the two cases.
The Averaging Principle is often applied to differential equations after
they are transformed to one of the above standard forms. It consists of
approximating a solution of the angular standard form by the corresponding solution
(with the same initial conditions) of the averaged system
\[
\dot J=\epsilon\frac{1}{(2\pi)^n}\int_0^{2\pi}f(J,\theta)\, d\theta
\] 
or, for the time periodic standard form,
by the equation
\[
\dot y=\epsilon \frac{1}{T}\int_0^{T}F(y,t)\, dt.
\]
The validity of the approximation produced by the Averaging Principle is often left
unverified. However, as many examples show, the Averaging Principle is not
valid in general. Thus, a careful analysis of a physical problem must
always deal with this issue. Fortunately,  the problem has been treated in great 
detail in the mathematical literature. Many useful and rigorous
results are now available.

The basic result involves only one angular variable; that is, it pertains
either to
the time periodic standard form, or to the case when $\theta$ is one dimensional
in the angular standard form.  The Averaging Theorem asserts in this
case (with appropriate smoothness of the functions involved) 
that the averaged solution
approximates the original solution with an error of order  $\epsilon$ on a 
time-scale of order $1/\epsilon$; 
for example, if $x(0)=y(0)$, then there are constants
$C_1$ and $C_2$ that do not depend on $\epsilon$ such that
\[ |x(t)-y(t)|<C_1\epsilon, \qquad 0\le t\le \frac{C_2}{\epsilon}.\]
Furthermore, there is a change of coordinates (the averaging
transformation) of the form
$x=z+\epsilon K(z,t)$ such that
\begin{equation}\label{B2}
\dot z= \epsilon \frac{1}{T}\int_0^{T}F(z,t)\, dt+\epsilon^2 F_2(z,t)+
\epsilon^3{\cal F}(z,t,\epsilon).
\end{equation}
This is exactly the idea that we use to obtain the averaging 
transformation~(\ref{avertran}): The purpose of the averaging
transformation is to render the transformed equation in a form that is already 
averaged to first order.
The second order averaged system
is then 
\begin{equation}\label{B3}
\dot u=\epsilon \frac{1}{T}\int_0^{T}F(u,t)\, dt
   +\epsilon^2 \frac{1}{T}\int_0^TF_2(u,t)\,dt.
\end{equation}
That is, once the differential equation~(\ref{tpsf}) is transformed into
the exact form~(\ref{B2}) using an averaging transformation, the second
order averaged system is simply obtained from~(\ref{B2}) by
averaging the second term and dropping terms of higher order. 
The averaging transformation can be uniquely defined as in Section~2;
indeed, this leads to a unique second order averaged system.
It turns out that a solution $u(t,u_0,\epsilon)$ of this system with initial condition 
$u(0,u_0,\epsilon)=u_0$ can be used to approximate the corresponding solution
of equation~(\ref{tpsf}) with an error of order $\epsilon^2$ on a time-scale
of order $1/\epsilon$. More precisely, there are constants $c_1>0$ and $c_2>0$
such that 
\[
|x(t,u_0+\epsilon K(u_0,0),\epsilon)-
   [u(t,u_0,\epsilon)+\epsilon K(u(t,u_0,\epsilon),t)]|<c_1\epsilon^2
\]
for $0\le t\le c_2/\epsilon$.
Thus, the second order averaged system
gives a second order approximation to the full dynamics.
We expect that the qualitative features of the dynamics of the
original system, for example, the stability
of its rest points, are reflected in the second order averaged system.
While this is not always the case, it is a reasonable working
hypothesis for the analysis of our model. 
The {\em partially} averaged system is the result of the application of the
method of averaging to the dynamical system at {\em resonance}.
Some of the delicate mathematical issues associated with
higher order averaging, together with an
analysis of resonance trapping for a one degree of freedom model, are
discussed in~\cite{murdock1}.

It is important to note that for systems in angular standard form with more than
one angle the Averaging Principle is not valid in general. In fact, it is exactly
the orbits for which the Averaging Principle is invalid that correspond to
the orbits that are captured into resonances. In~\cite{cmr2}, we have discussed
the implications of the result of Neishtadt, which asserts that, under additional
assumptions that hold for our model, 
the set of initial conditions corresponding to capture has
measure of order $\epsilon^{1/2}$. Resonance trapping is therefore proved
to be rare for the case of two angles unless Neishtadt's hypotheses are violated
(see~\cite{bj} for a discussion).
Rigorous results about the validity of the averaging method
for the case of more than two angles are much weaker than Neishtadt's result 
(cf.~\cite{arnold2}). 
%\begin{thebibliography}{99}

\begin{verbatim}
To obtain the figures for this paper please send a request to
carmen@chicone.math.missouri.edu
\end{verbatim}
\begin{figure}[h]
\vspace*{2in}
\caption[]{ 
The plots are for for system~(\ref{MathEQM}) with parameter values
$\epsilon=10^{-4}$, $\delta/\epsilon=10^{-3}$,
$\alpha=1$, $\beta=0$, $\rho=0$, and $\Omega= 1$. 
The top panel shows $L=a^{1/2}$ versus time for the initial conditions
$(p_r,p_\theta,r,\theta)$
equal to $(0.2817, 0.6389, 1.6628, 2.9272)$.  The middle
panel shows eccentricity versus time and the bottom panel shows $G$ versus $L$.
Here, $L_*=1$ corresponds to $(1:1)$ resonance,
$1/L_*^3=\Omega$.
\label{fig1}}
\end{figure}
\begin{figure}[h]
\caption[]{
The plots are for system~(\ref{MathEQM}) with parameter values
$\epsilon=10^{-4}$, $\delta/\epsilon=10^{-3}$,
$\alpha=1$, $\beta=0$, $\rho=0$, and $\Omega= 1$.
The top panel shows $L=a^{1/2}$ versus time for the initial conditions
$(p_r,p_\theta,r,\theta)$
equal to $(-0.007485536,   0.850305,    2.758946,   0.97661369282041)$.  The middle
panel shows eccentricity versus time and the bottom panel shows $G$ versus $L$.
Here, $L_*\approx 1.2599$ corresponds to $(2:1)$ resonance,
$2/L_*^3=\Omega$.
\label{fig2}}
\end{figure}
\begin{figure}[h]
\caption[]{
The plots are for system~(\ref{MathEQM}) with parameter values
$\epsilon=10^{-4}$, $\delta/\epsilon=10^{-3}$,
$\alpha=1$, $\beta=0$, $\rho=0$, and $\Omega= 1$.
The top panel shows $L=a^{1/2}$ versus time for the initial conditions
$(p_r,p_\theta,r,\theta)$
equal to $(-.09951, 1.56904, 2.33813, 2.32909)$.  The middle
panel shows eccentricity versus time and the bottom panel shows $G$ versus $L$.
Here, $L_*\approx 1.5874$ corresponds to $(4:1)$ resonance,
$4/L_*^3=\Omega$.
\label{fig3}}
\end{figure}

\begin{references}
\bibitem{cmr} Chicone C, Mashhoon B and  Retzloff D G 1996 Gravitational
ionization: periodic orbits of binary systems perturbed by gravitational
radiation {\em Annales de l'Institut Henri Poincar\'{e}, Physique Th\'{e}orique}
{\bf 64} 87--125
\bibitem{cmr1} Chicone C, Mashhoon B and  Retzloff D G 1996 On the
ionization of a Keplerian binary system by periodic gravitational radiation
{\em J. Math.\ Phys.} {\bf 37} 3997--4016
\bibitem{cmr2} Chicone C, Mashhoon B and  Retzloff D G 1996 Gravitational
ionization: a chaotic net in the Kepler system,
preprint
\bibitem{hulse} Hulse R A and Taylor J H 1975
Discovery of a pulsar in a binary system {\em Astrophys.\ J.} {\bf 195} L51--L53
\bibitem{taylor} Taylor J H, Wolszczan A, Damour T and Weisberg J M 1992
Experimental constraints on strong-field relativistic gravity
{\em Nature} {\bf 355} 132--136
\bibitem{melita} Melita M D and Woolfson M M 1996 Planetary
commensurabilities driven by accretion and dynamical friction, {\em Mon. Not. Roy.
Astron. Soc.} {\bf 280} 854-862
\bibitem{Kov} Kovalevsky J 1967 {\em Introduction to Celestial Mechanics,}
{\em Astrophysics and Space Science Library Vol.~ 7} (New York: Springer-Verlag)
\bibitem{Stern} Sternberg S 1969 {\em Celestial Mechanics (Vol 1--2)}
(New York:  Benjamin)
\bibitem{poincare} Poincar\'{e} H  1892-99
{\em Les M\'{e}thodes Nouvelles de la M\'{e}canique C\'{e}leste (Vol 1-3)}
(Paris: Gauthier-Villars)
\bibitem{mashoon1}  Mashhoon B 1978 On tidal resonance
{\em Astrophys.\ J.}
{\bf 223} 285--298; 1977 Tidal radiation  {\em Astrophys.\ J.} {\bf 216} 591--609;
1979 On the detection of gravitational
radiation by the Doppler tracking of spacecraft {\em Astrophys.\ J.}
{\bf 227} 1019--1036; 1980 Absorption of gravitational radiation
{\em GRG Abstracts} (Jena Meeting), 1505--1507; 
Mashhoon B, Carr B J and Hu B L 1981
The influence of cosmological gravitational waves on a Newtonian binary system
{\em Astrophys.\ J.} {\bf 246} 569--591
\bibitem{ccc} Chicone C 1995  A geometric approach to regular perturbation
theory with an application to hydrodynamics
{\em Trans.\  Amer.\ Math.\ Soc.} {\bf 347} 4559--4598
\bibitem{henrard} Henrard J 1985 Resonance sweeping in the solar system
{\em Stability of the Solar System and Its Minor Natural and Artificial Bodies}
Szebehely V G (ed) (Dordrecht: Reidel) 183--192
\bibitem{beauge} Beaug\'{e} C, Aarseth S J and Ferraz-Mello S 1994 Resonance
capture and the formation of the outer planets {\em Mon. Not. Roy. Astron. Soc.}
{\bf 270} 21-34
\bibitem{gomes} Gomes R S 1995 The effect of nonconservative forces on
resonance lock: stability and instability {\em Icarus} {\bf 115} 47-59
\bibitem{lai} Lai D 1996 Orbital decay of the PSR J0045-7319 binary system:
age of radio pulsar and initial spin of neutron star
{\em Astrophys. J.} {\bf 466} L35--L38
\bibitem{arnold2} Arnold V I (ed) 1988 {\em Dynamical Systems III},
{\em Encyclopedia of Mathematical Sciences  (Vol 3)} (New York: Springer-Verlag)
\bibitem{bj} Burns  T J Jones C K R T  1993 A mechanism for capture into resonance
{\em Physica D} {\bf 69}  85--106
\bibitem{wig} Wiggins S 1990 {\em Introduction to Applied Nonlinear Dynamical
Systems and Chaos} (New York: Springer-Verlag)
\bibitem{sv} Sander J A Verhulst F 1985 {\em Averaging Methods in
Nonlinear Dynamical Systems} (New York: Springer-Verlag)
\bibitem{watson} Watson G N 1966 {\em A Treatise on the Theory of Bessel
Functions} (New York: Cambridge University Press)
\bibitem{murdock} Murdock J A 1991 {\em Perturbations: Theory and Methods}
(New York: Wiley)
\bibitem{chirikov} Chirikov B V 1979 A universal instability of many-dimensional
oscillator systems {\em Phys. Rep.} {\bf 52} 262--379
\bibitem{murdock1} Murdock J A 1988 Qualitative theory of nonlinear resonances by
averaging and dynamical systems methods {\em Dynamics Reported Vol.~1}
(New York: Wiley)
%\end{thebibliography}
\end{references}
\end{document}